\documentclass[10pt,twocolumn,letterpaper]{article}

\usepackage[pagenumbers]{cvpr} % To force page numbers, e.g. for an arXiv version

\usepackage{bm}
\usepackage{amsmath,graphicx}
\usepackage{amssymb}
\usepackage{makecell}
\usepackage{multirow}
\usepackage{multicol}
\usepackage{booktabs}
\usepackage{algorithm,algpseudocode}

\usepackage{tipa}
\usepackage{siunitx}
\usepackage{textcomp}
\usepackage{soul}
\usepackage{xcolor}
\usepackage{cite}
\usepackage{pifont}
\newcommand{\cmark}{\ding{51}}
\newcommand{\xmark}{\ding{55}}
\usepackage{lipsum}
\usepackage{arydshln}

\usepackage[pagebackref,breaklinks,colorlinks]{hyperref}

\usepackage[capitalize]{cleveref}
\crefname{section}{Sec.}{Secs.}
\Crefname{section}{Section}{Sections}
\Crefname{table}{Table}{Tables}
\crefname{table}{Tab.}{Tabs.}

\def\cvprPaperID{5013} % *** Enter the CVPR Paper ID here
\def\confName{CVPR}
\def\confYear{2023}

\begin{document}

%%%%%%%%% TITLE - PLEASE UPDATE
% \title{Watch or Listen: Robust Audio-Visual Speech Recognition\\via multimodal Stream Reliability Scoring}
\title{Watch or Listen: Robust Audio-Visual Speech Recognition\\with Visual Corruption Modeling and Reliability Scoring}
\author{Joanna Hong\thanks{Both authors have contributed equally to this work.} \quad\quad Minsu Kim\footnotemark[1] \quad\quad Jeongsoo Choi \quad\quad Yong Man Ro\thanks{Corresponding author} \\
Image and Video Systems Lab, KAIST\\
{\tt\small \{joanna2587, ms.k, jeongsoo.choi, ymro\}@kaist.ac.kr}
% For a paper whose authors are all at the same institution,
% omit the following lines up until the closing ``}''.
% Additional authors and addresses can be added with ``\and'',
% just like the second author.
% To save space, use either the email address or home page, not both
}
\maketitle

%%%%%%%%% ABSTRACT
\begin{abstract}
This paper deals with Audio-Visual Speech Recognition (AVSR) under multimodal input corruption situations where audio inputs and visual inputs are both corrupted, which is not well addressed in previous research directions. Previous studies have focused on how to complement the corrupted audio inputs with the clean visual inputs with the assumption of the availability of clean visual inputs. However, in real life, clean visual inputs are not always accessible and can even be corrupted by occluded lip regions or noises. Thus, we firstly analyze that the previous AVSR models are not indeed robust to the corruption of multimodal input streams, the audio and the visual inputs, compared to uni-modal models. Then, we design multimodal input corruption modeling to develop robust AVSR models. Lastly, we propose a novel AVSR framework, namely Audio-Visual Reliability Scoring module (AV-RelScore), that is robust to the corrupted multimodal inputs. The AV-RelScore can determine which input modal stream is reliable or not for the prediction and also can exploit the more reliable streams in prediction. The effectiveness of the proposed method is evaluated with comprehensive experiments on popular benchmark databases, LRS2 and LRS3. We also show that the reliability scores obtained by AV-RelScore well reflect the degree of corruption and make the proposed model focus on the reliable multimodal representations.
\end{abstract}

\vspace{-0.4cm}
%%%%%%%%% BODY TEXT
\section{Introduction}
\vspace{-0.1cm}
\label{sec:intro}
Imagine you are watching the news on Youtube. Whether the recording microphone is a problem or the video encoding is wrong, the anchor's voice keeps breaking off, so you cannot hear well. You try to understand her by her lip motions, but making matters worse, the microphone keeps covering her mouth, so the news is hardly recognizable. These days, people often face these kinds of situations, even in video conferences or interviews where the internet situation cuts in and out.

As understanding speech is the core part of human communication, there have been a number of works on speech recognition \cite{amodei2016deep2,assael2016lipnet}, especially based on deep learning. These works have tried to enhance audio representation for recognizing speech in a noisy situation \cite{weninger2015speech, tan2019learning, wang2020complex, braun2021towards} or to utilize additional visual information for obtaining complementary effects \cite{noda2015audio, afouras2018deep, vaswani2017attention, petridis2018end, gulati2020conformer, hong2022vcafe}. Recently, even technologies that comprehend speech from only visual information have been developed \cite{wang2019multigrained, xiao2020deformation, zhao2020mi, zhang2020cutout, Kim_2021_ICCV, kim2021lip, kim2023lip, hong2021speech, hong2022visagesyntalk}. 

With the research efforts, automatic speech recognition technologies including Audio Speech Recognition (ASR), Visual Speech Recognition (VSR), and Audio-Visual Speech Recognition (AVSR) are achieving great developments with outstanding performances \cite{baevski2020wav2vec2,kim2022distinguishing,ma2021end}. With the advantages of utilizing multimodal inputs, audio and visual, AVSR that can robustly recognize speech even in a noisy environment, such as in a crowded restaurant, is rising for the future speech recognition technology. However, the previous studies have mostly considered the case where the audio inputs are corrupted and utilizing the additional clean visual inputs for complementing the corrupted audio information. Looking at the case, we come up with an important question, \textit{what if both visual and audio information are corrupted, even simultaneously?} In real life, like the aforementioned news situation, cases where both visual and audio inputs are corrupted alternatively or even simultaneously, are frequently happening. To deal with the question, we firstly analyze the robustness of the previous ASR, VSR, and AVSR models on three different input corruption situations, 1) audio input corruption, 2) visual input corruption, and 3) audio-visual input corruption. Then, we show that the previous AVSR models are not indeed robust to audio-visual input corruption and show even worse performances than uni-modal models, which is eventually losing the benefit of utilizing multimodal inputs.

To maximize the superiority of using multimodal systems over the uni-modal system, in this paper, we propose a novel multimodal corruption modeling method and show its importance in developing robust AVSR technologies for diverse input corruption situations including audio-visual corruption. To this end, we model the visual corruption with lip occlusion and noises that are composed of blurry frames and additive noise perturbation, along with the audio corruption modeling. Then, we propose a novel AVSR framework, namely Audio-Visual Reliability Scoring module (AV-RelScore), that can evaluate which modal of the current input representations is more reliable than others. The proposed AV-RelScore produces the reliability scores for each time step, which represent how much the current audio features and the visual features are helpful for recognizing speech. With the reliability scores, meaningful speech representations can be emphasized at each modal stream. Then, through multimodal attentive encoder, the emphasized multimodal representations are fused by considering inter-modal relationships. Therefore, with the AV-RelScore, the AVSR model can refer to the audio stream when the given visual stream is determined as less reliable (\ie, corrupted), and vice versa. We provide the audio-visual corruption modeling for the reproducibility and the future research.\footnote{\url{https://github.com/joannahong/AV-RelScore}}

Our key contributions are as follows:
\vspace{-0.2cm}
\begin{itemize}
\item To the best of our knowledge, this is the first attempt to analyze the robustness of deep learning-based AVSR under the corruption of multimodal inputs including lip occlusions.
\vspace{-0.2cm}
\item We propose an audio-visual corruption modeling method and show that it is key for developing robust AVSR technologies under diverse environments.
\vspace{-0.2cm}
\item We propose Audio-Visual Reliability Scoring module (AV-RelScore) to figure out whether the current input modal is reliable or not, so that to robustly recognize the input speech even if one modality is corrupted, or even both.
\vspace{-0.2cm}
\item We conduct comprehensive experiments with ASR, VSR, and AVSR models to validate the effectiveness of the proposed audio-visual corruption modeling and AV-RelScore on LRS2 \cite{chung2017lrs2} and LRS3 \cite{afouras2018lrs3}, the largest audio-visual datasets obtained in the wild.
\end{itemize}  

%------------------------------------------------------------------------
\section{Related Works}
\label{sec:relatedworks}
\subsection{Audio-visual speech recognition}
There has been a great development in automatic speech recognition in both audio-based (ASR) and visual-based (VSR), along with the progress of deep learning. 

%------------------------------------ Figure 1
%#################################################
\begin{figure}[t]
	\begin{minipage}[b]{1.0\linewidth}
		\centering
		\centerline{\includegraphics[width=7.3cm]{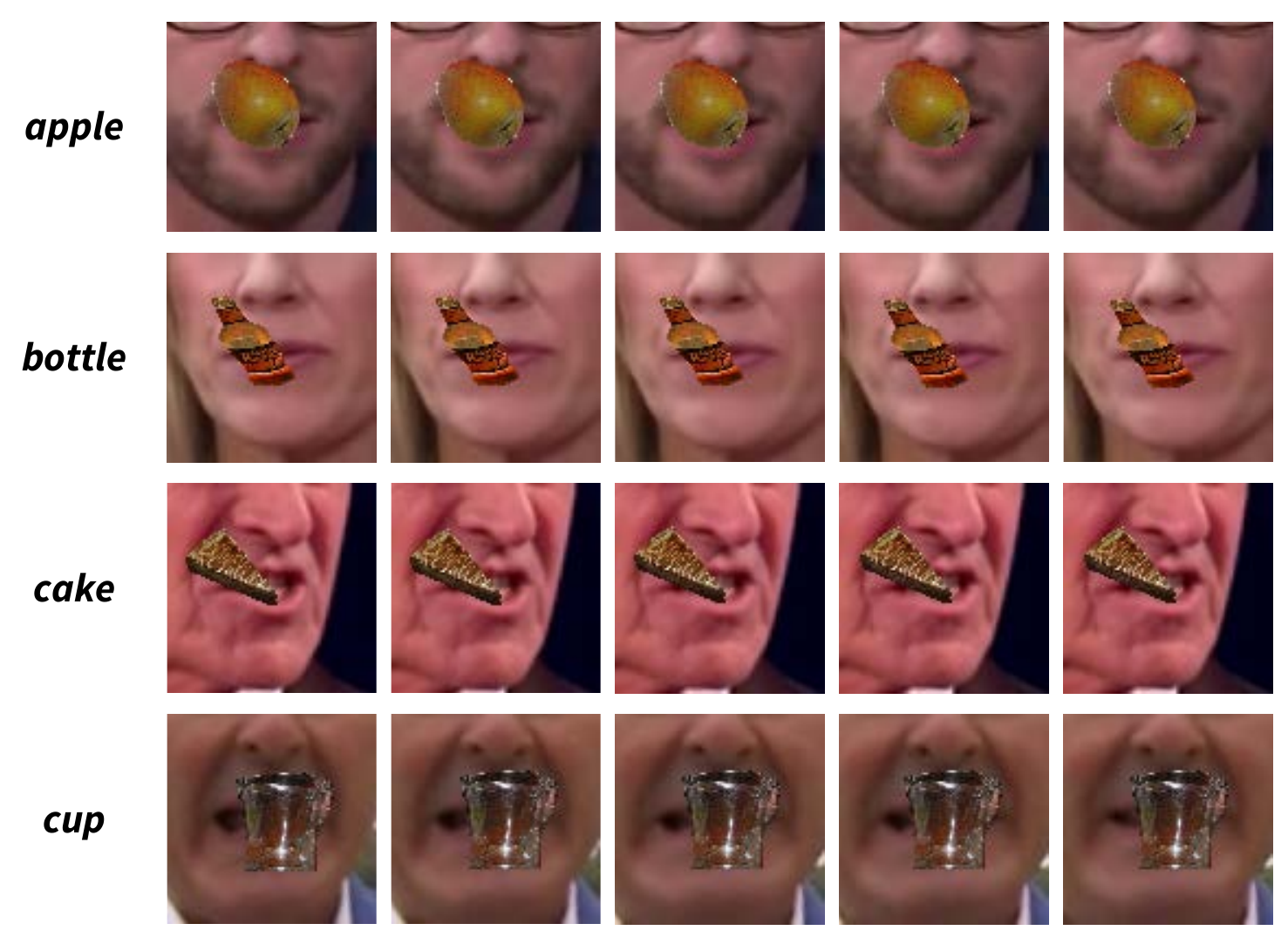}}
	\end{minipage}
 	\vspace{-0.6cm}
	\caption{Examples of visual occlusion with NatOcc patches.}
	\label{fig:1}
	\vspace{-0.2cm}
\end{figure}
%##################################################

%------------------------------------ Figure 2
%#################################################
\begin{figure}[t]
	\begin{minipage}[b]{1.0\linewidth}
		\centering
		\centerline{\includegraphics[width=7.3cm]{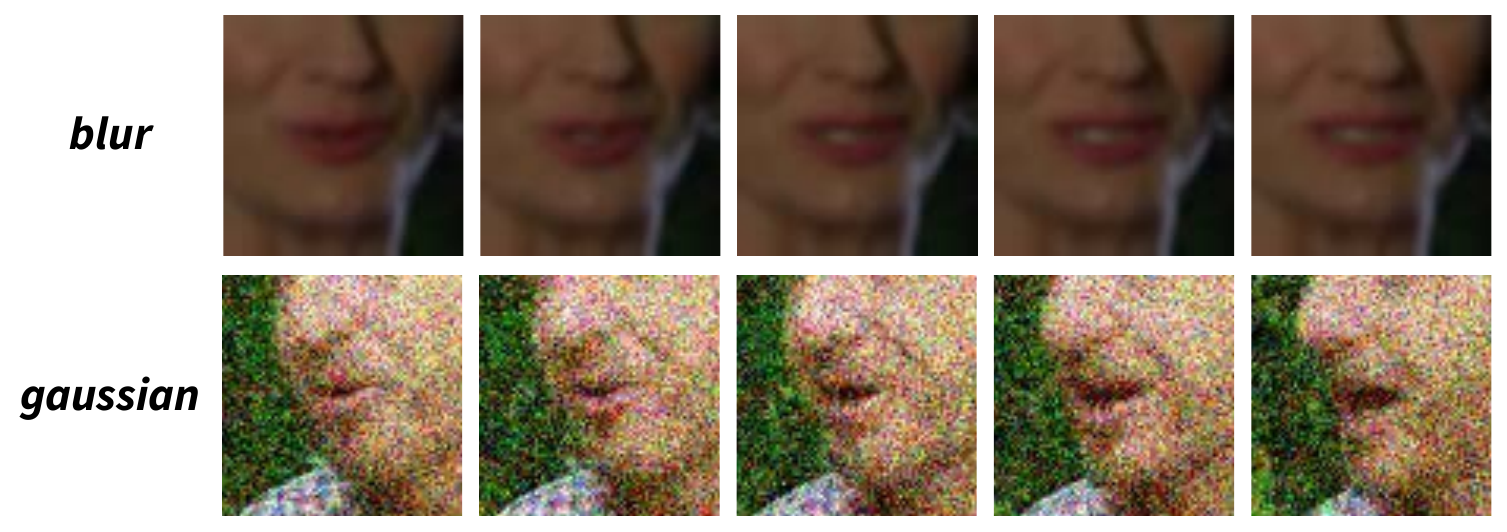}}
	\end{minipage}
  	\vspace{-0.6cm}
	\caption{Examples of visual corruption with noises.}
   	\vspace{-0.4cm}
	\label{fig:2}
\end{figure}
%##################################################

%------------------------------------ Figure 3
%#################################################
\begin{figure*}[t]
	\begin{minipage}[b]{1.0\linewidth}
		\centering
		\centerline{\includegraphics[width=17.3cm]{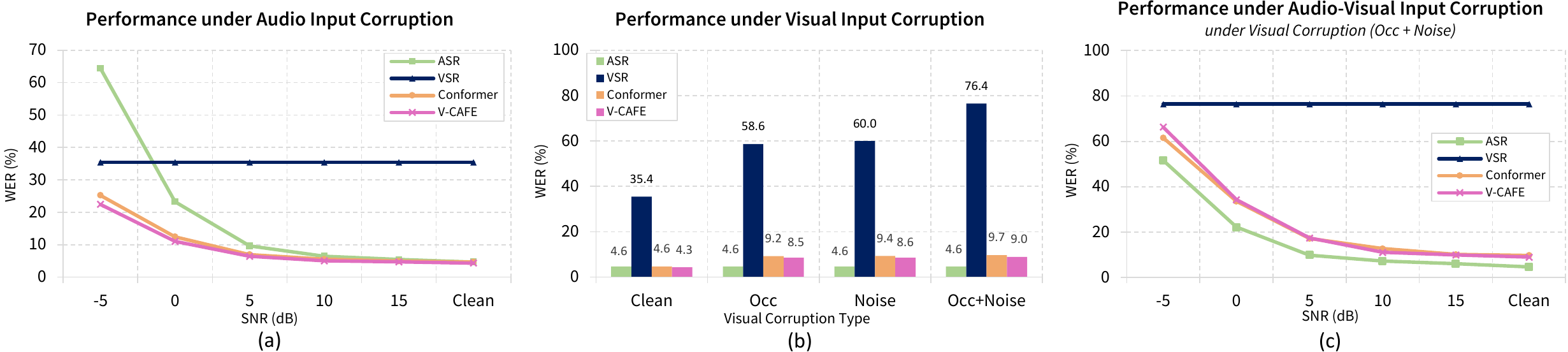}}
	\end{minipage}
 	\vspace{-0.7cm}
	\caption{Speech recognition performances of ASR, VSR, and AVSR models on LRS2 dataset under different input corruption types: (a) Audio input corruption with babble noise. (b) Visual input corruption with occlusion and noise. (c) Audio-visual input corruption.}
	\label{fig:3}
	\vspace{-0.4cm}
\end{figure*}
%##################################################

Deep Neural Networks (DNNs) were embedded in the standard ASR pipeline \cite{mohamed2011acoustic,hinton2012speech,dahl2011context,seide2011conversational}. Convolutional neural networks \cite{abdel2012applying,abdel2014convolutional,sainath2013improvements} and recurrent neural networks \cite{graves2013speech,graves2014towards,sak2014long} brought large improvement in ASR performances. Connectionist Temporal Classification (CTC) \cite{graves2006ctc} and Sequence-to-Sequence \cite{sutskever2014sequence} learning methods were developed for end-to-end speech recognition. With advanced DNN architectures such as Transformer \cite{vaswani2017attention} and self-supervised/unsupervised learning methods, recent ASR methods \cite{hannun2014deep,amodei2016deep2,baevski2020wav2vec2,schneider2019wav2vec,baevski2019vq,gulati2020conformer} achieved significant performances sufficient to be used in the real world. With the demands on recognizing speech even if there is no speech audio available, VSR has been developed mainly based on the developed technology in ASR. \cite{chung2017lrs2,petridis2018end,kim2021cromm,kim2022distinguishing,martinez2020lipreading,zhang2019spatio,zhao2020hearing,ma2022visual,kim2022speaker,kim2023prompt, prajwal2022sub} narrowed the performance gap between ASR and VSR by proposing network architecture and learning methods specialized for VSR.

By combining the two technologies, Audio-Visual Speech Recognition (AVSR) is developed. Since AVSR is robust to acoustic noises by complementing the corrupted information with visual inputs, it is preferred under wild environments and real-world applications. With the pioneers of AVSR models \cite{huang2013audio, mroueh2015deep, noda2015audio, stewart2013robust}, the research of AVSR has been placed in the mainstream of the deep multimodal field. Recent works have shown much greater improvements in AVSR \cite{yu2020audio,petridis2018audio}; deep AVSR with large-scale datasets \cite{afouras2018deep}, a two-stage speech recognition model \cite{xu2020discriminative} that firstly enhances the speech with visual information then performs recognition, AVSR using Conformer architecture \cite{ma2021end}, and speech enhanced AVSR \cite{hong2022vcafe} were proposed. While there have been great developments in AVSR, there was a lack of consideration for visual impairment in AVSR. In this paper, we analyze that the previous deep learning-based AVSR models are weak for audio-visual input corruption and even show lower performance than ASR models. To maximize the advantages of using multimodal inputs, we propose 1) audio-visual corruption modeling to train robust AVSR models and 2) Audio-Visual Reliability Scoring module (AV-RelScore) to effectively utilize a more reliable input stream while suppressing unreliable representations.

\vspace{-0.13cm}
\subsection{Visual occlusion modeling}
\vspace{-0.2cm}
In the real world, we can occasionally meet some objects that are occluded by other objects. This makes the trained DNNs perform worse when they were trained without consideration of the occluded situation. For example, there are image classification \cite{cen2021deep,li2020yolo} where the target object is overlapped by other objects, and facial expression recognition \cite{li2018occlusion} where some facial parts are occluded by an object. To overcome this, many works tried to model the occlusion during training so that the trained network can robustly perform on its given task \cite{cen2020classification,li2018patch,voo2022delving,wang2020robust}. We try to model the lip occluded situation that frequently occurs when the speaker uses a mic or eats some food by using Naturalistic Occlusion Generation (NatOcc) of \cite{voo2022delving}. We show that visual corruption modeling is as important as audio corruption modeling, and is important to build a robust AVSR model.

%------------------------------------------------------------------------
\vspace{-0.13cm}
\section{Methodology}
\vspace{-0.1cm}
\label{sec:Methodlogy}
Let $\bm{x}_v \in \mathbb{R}^{T \times H \times W \times C}$ be a lip-centered talking face video with frame lengths of $T$ and frame size of $H \times W \times C$, and $\bm{x}_a \in \mathbb{R}^{S}$ be a paired speech audio with the input video, where $S$ represents the length of audio. The objective of AVSR model is translating audio-visual inputs into ground-truth text, $y$. Ideally, since the model utilizes two input modalities, it also can robustly perform the recognition when one modal input is corrupted with environmental noise by leaning on the other modality. However, previous AVSR models failed to consider the corruption of visual inputs and only considered acoustic corruption. In the following subsections, we firstly analyze the robustness of the previous AVSR models on diverse environments, and present a robust AVSR method for both acoustic and visual corruption.

\subsection{Robustness of AVSR to acoustic and visual noise} 
\vspace{-0.15cm}
%pretrained model without trained with visual corruption.
In this subsection, we analyze the robustness of the previous AVSR models \cite{ma2021end,hong2022vcafe}, an ASR model\cite{hong2022vcafe}, and a VSR model\cite{ma2021end} on three different corrupted input situations, 1) audio input corruption, 2) visual input corruption, and 3) audio-visual input corruption, using LRS2 dataset. We directly utilize pre-trained models that do not consider the visual input corruption during training, in order to analyze their performances in different situations. For the audio input corruption, we injected a babble noise of \cite{varga1993noisex92} to the entire audio with different Signal-to-Noise Ratio (SNR) levels, -5 to 15dB, following \cite{ma2021end,hong2022vcafe}. For the visual input corruption, there can be various noise types, additive noise, blur, color distortion, occlusion, etc. We investigate two different types of visual corruption, occlusion and noise (blur + additive noise), that we can frequently face in practice. For the audio-visual input corruption, both audio corruption and visual corruption are injected for random chunks of each stream so that both streams can be corrupted simultaneously or alternatively.

\vspace{0.08cm}
\noindent {\bf Audio input corruption.} Since audio and visual inputs are highly correlated for the speech content instead of the background noise, AVSR models are expected to robustly recognize speech compared to the ASR model. As shown in Figure \ref{fig:3}(a), the performance of ASR, Word Error Rate (WER), is steeply degraded when the noise level becomes higher (\ie, lower SNR). On the other hand, AVSR models, Conformer \cite{ma2021end} and V-CAFE \cite{hong2022vcafe}, show the robustness against acoustic noise by complementing the noisy audio signals with the visual inputs. Since the VSR model is not affected by the acoustic noise, it shows consistent performance for all SNR ranges. This result is in line with our expectations that AVSR models would be robust against acoustic noises.

\vspace{0.08cm}
\noindent {\bf Visual input corruption.} Similar to audio input corruption, we perturb the visual inputs and examine performances of each model. Figure \ref{fig:3}(b) shows the results under three different types of visual corruption, occlusion, noise, and both occlusion and noise. The performance of VSR model is largely affected by the visual corruption and nearly crushed when both occlusion and noises are applied to the input. The AVSR models also show degraded performances so that Conformer loses 5.1\% WER and V-CAFE loses 4.7\% WER, from their original performances (\ie, clean situation), when both visual corruptions are applied. We found that the AVSR models tend to depend on the audio stream and are somewhat robust to visual input-only corruptions compared to VSR. Since ASR model is not affected by visual corruption and shows the best performance, the results indicate that it would be better to use the ASR model instead of the AVSR model when we know the inferring environment has clean audio and perturbed visual inputs.

\vspace{0.08cm}
\noindent {\bf Audio-visual input corruption.} Finally, we model audio-visual input corruption where both audio and visual sequences are randomly corrupted, to confirm the robustness of previous models against multimodal corruption. The cases of alternative corruption of audio and visual sequences and simultaneous corruption are included. Ideally, when audio is corrupted, the model is expected to utilize the visual stream to produce appropriate results and vice versa. The results on different audio SNR levels with visual corruption using both occlusion and noise are illustrated in Figure \ref{fig:3}(c). In this situation, the previous AVSR models, Conformer and V-CAFE, show even worse performances than the audio-only model (\ie, ASR) on all SNR ranges. As the previous AVSR models are largely depending on the audio stream and they tend to refer to the visual stream when the audio input is corrupted, if both audio and visual inputs are corrupted, they fail in complementing from the multimodal information. Especially, V-CAFE that explicitly enhances the corrupted audio with visual inputs shows the worst performance than Conformer that doesn't contain the audio enhancement part, under strong noise situations (-5 to 5dB SNR). This shows the risk when only considering the audio corruption during training and model designing; when there comes a visually perturbed input, the trained model's performance can be degraded. Therefore, with the currently developed AVSR models, when there is audio-visual corruption, it would be better to use the audio-only model (\ie, ASR) instead of the multimodal model (\ie, AVSR).

From the examples of three input corruption cases, we show that even if we use two modal inputs, audio and video, in AVSR, AVSR model is not always robust to different input corruption cases. Therefore, to develop a robust AVSR method for maximizing the advantages of using multimodal inputs, we should model the visual input corruption case as well as the audio input corruption during training, and design the appropriate model architecture. In the following subsection, we introduce visual input corruption modeling for developing a robust AVSR model.

%------------------------------------ Figure 4
%#################################################
\begin{figure*}[t]
	\begin{minipage}[b]{1.0\linewidth}
		\centering
		\centerline{\includegraphics[width=15.0cm]{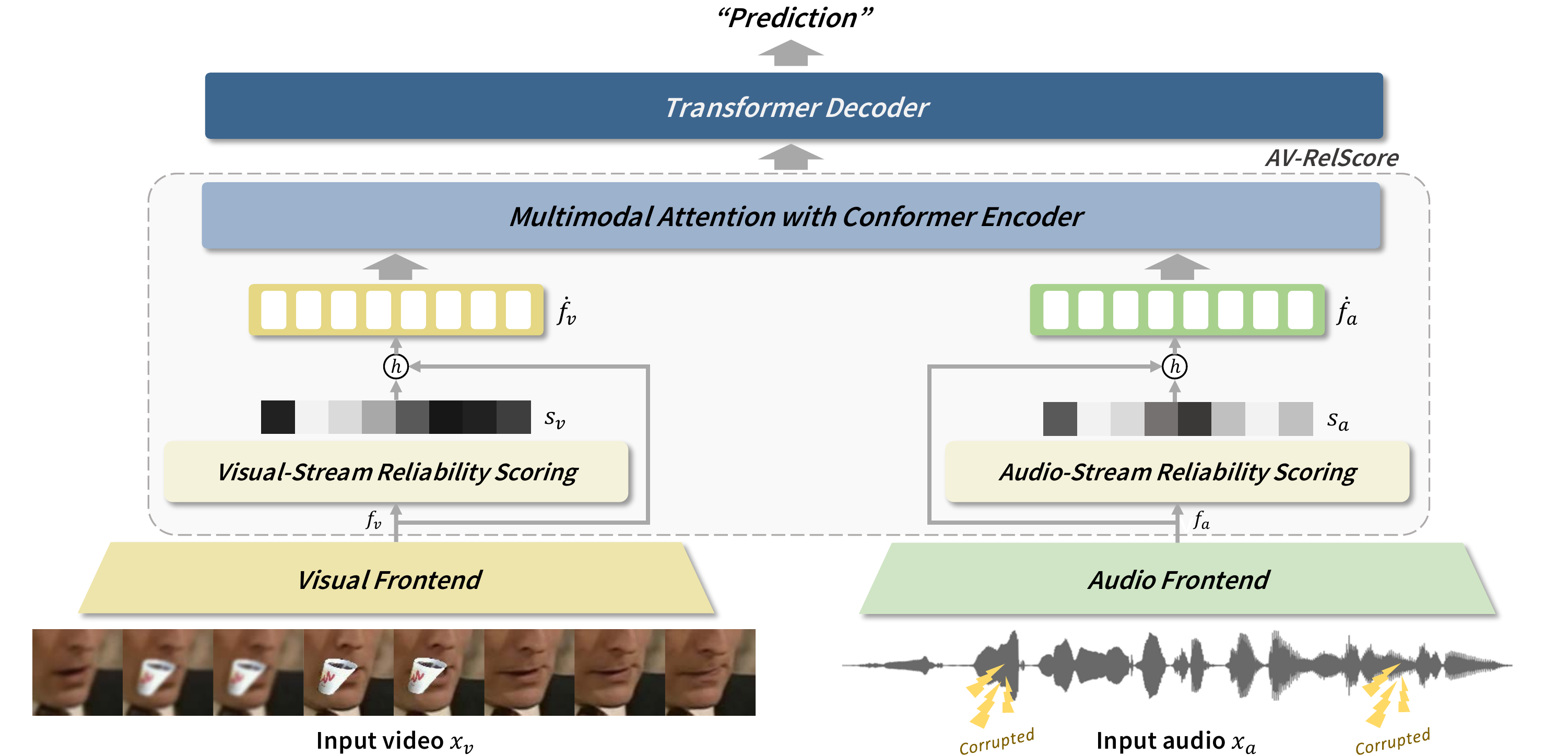}}
	\end{minipage}
	\caption{Overall architecture of the proposed AVSR framework.}
	\label{fig:4}
	\vspace{-0.3cm}
\end{figure*}
%##################################################

% \vspace{-0.1cm}
\subsection{Visual corruption modeling}
\vspace{-0.15cm}
Acoustic noise modeling is well-known and generally utilized in both ASR and AVSR; it can be modeled by injecting various environmental noise data \cite{thiemann2013demand,varga1993noisex92} into clean audio inputs, rejecting some frequency ranges, and distorting the signals \cite{ko2015audio}. However, previous works have missed considering visual noise modeling, which is important for building robust AVSR models. We proposed to model the visual input corruption with occlusion patches and noises.

\vspace{-0.35cm}
\subsubsection{Visual corruption with occlusion patch}
\vspace{-0.2cm}
The occlusion is frequently induced when a speaker gives a speech with a microphone or a script. The lips of the speaker can be repeatedly occluded by such objects, thus hard to recognize speech by solely watching the lip movements. To simulate the occlusion, we introduce attaching patches to the region of lips. We utilize Naturalistic Occlusion Generation (NatOcc) patches from high-quality synthetic face occlusion datasets of \cite{voo2022delving} that are originally designed for occlusion-aware face segmentation tasks. The NatOcc patches consist of various objects generally seen in our everyday lives, such as fruits, desserts, cups, and so on. Since it is designed for producing naturalistic occluded faces, it is appropriate for our lip occlusion modeling.

Given the input lip-centered talking face video $\bm{x}_v$ with length $T$, we randomly choose, $N$, representing how many times the occluded chunk occurs in whole sequences. We design that the patches are not overlapped throughout the entire input video. To do so, the input video frames $\bm{x}_v$ are evenly divided by occurrence number $N$, called segment lip videos $\bm{x}_v = \{x_{v,n}\}_{n=1}^N$. 
Then, we select a random ratio $t$ with the range in $0.3$ to $0.5$ of the segment of video frames, $x_{v,n}$, where we are going to attach the patch. Then the patch is located on the random position of lip landmarks. We insert the occlusion patch with a probability of $0.8$ and otherwise we use clean visual inputs. The examples of visual corruption with occlusion are shown in Figure \ref{fig:1}.

\vspace{-0.35cm}
\subsubsection{Visual corruption with noise}
\vspace{-0.2cm}
Furthermore, visual corruption can occur when there are actual noises in the input video sequences. These kinds of problems usually occur when the encoding process goes wrong, the camera is out of focus, or there are communication issues. To implement these noise-corrupted video situations, we adopt two types of noises: blur and additive noise. We randomly insert blur or Gaussian noise to input face video $\bm{x}_v$ with a probability of $0.3$, respectively; otherwise, we utilize the clean sequence. For visual corruption with noise, we also follow the same scheme of selecting the random ratio $t$ and occurrence number $N$ applied in the visual corruption with occlusion patches. The examples of resulting noise corruption are indicated in Figure \ref{fig:2}. By combining the two visual corruption, occlusion and noises, we can train an AVSR model with visual input corruption along with audio input corruption.

% \vspace{-0.1cm}
\subsection{Audio-visual reliability scoring}
\vspace{-0.2cm}
In addition to the audio-visual corruption modeling, we propose a novel AVSR model that can robustly recognize speech under diverse noise situations. 

Humans can easily be aware of whether the given speech audio is noisy or clean just by hearing. Similarly, corrupted video such as occlusion or blur in the lip region is also easily detectable just by watching the video. In analogy to the human input systems, we propose Audio-Visual Reliability Scoring module (AV-RelScore) which can determine whether the input audio or video is corrupted or not, and minimize the effect of the corrupted stream in prediction. Therefore, the model can focus more on the other modal stream for speech modeling when one modal is determined as corrupted. Also, if it is determined as both modalities are corrupted, the model can pay attention more to capture context to infer the corrupted speech. The overall architecture of the proposed AVSR framework is illustrated in Figure \ref{fig:4}.

Each modal feature, audio feature $f_a\in\mathbb{R}^{T\times D}$ and visual feature $f_v\in\mathbb{R}^{T\times D}$, is embedded through modality-specific front-ends, respectively. The audio front-end downsamples the time length of the input audio to have the same length as that of the visual feature (\ie, $T$). Then, to determine the corrupted frames, AV-RelScore is designed.
Firstly, Audio-Stream Reliability Scoring module inspects the audio features by modeling its temporal information with temporal convolutions, and outputs audio reliability scores $s_a\in\mathbb{R}^{T\times D}$ with the value range of $\left[0, 1\right]$, where 0 indicates less reliability of audio representation in speech modeling while 1 indicates full reliability. With the obtained reliability scores that inform which audio features are less knowledgeable for the speech modeling, we can emphasize the more reliable representations through the emphasis function, $h(a,b) = a + a \odot b$, where $\odot$ represents Hadamard product:
\begin{align}
    \dot{f}_a = h(f_a, s_a),
\end{align}
where $\dot{f}_a$ represents the emphasized audio features using the reliability score. Similar to the audio stream, we can obtain emphasized visual features from Visual-Stream Reliability Scoring module as follows,
\begin{align}
\dot{f}_v=h(f_v, s_v).
\end{align}
Therefore, the emphasized modality features, $\dot{f}_a$ and $\dot{f}_v$, respectively contain speech information of reliable frames while the corrupted representations are suppressed.

When frames of one modal features are determined as corrupted, it is important to refer to other modal features which possibly not corrupted. To this end, we encode the emphasized multimodal features, $\dot{f}_a$ and $\dot{f}_v$, using multimodal attentive encoder, so the inter-modal relationships can be considered. Inspired by \cite{chen2020uniter}, we concatenate multimodal features into the temporal dimension to create the combined emphasized modality features $\dot{f}_{av}=[\dot{f}_a, \dot{f}_v]\in\mathbb{R}^{2T\times D}$. Then, $\dot{f}_{av}$ is fed into attention-based network, Conformer \cite{gulati2020conformer}, to produce $\hat{f}_{av}\in\mathbb{R}^{T\times D}$.

\begin{figure}[t]
	\begin{minipage}[b]{1.0\linewidth}
		\centering
		\centerline{\includegraphics[width=8.1cm]{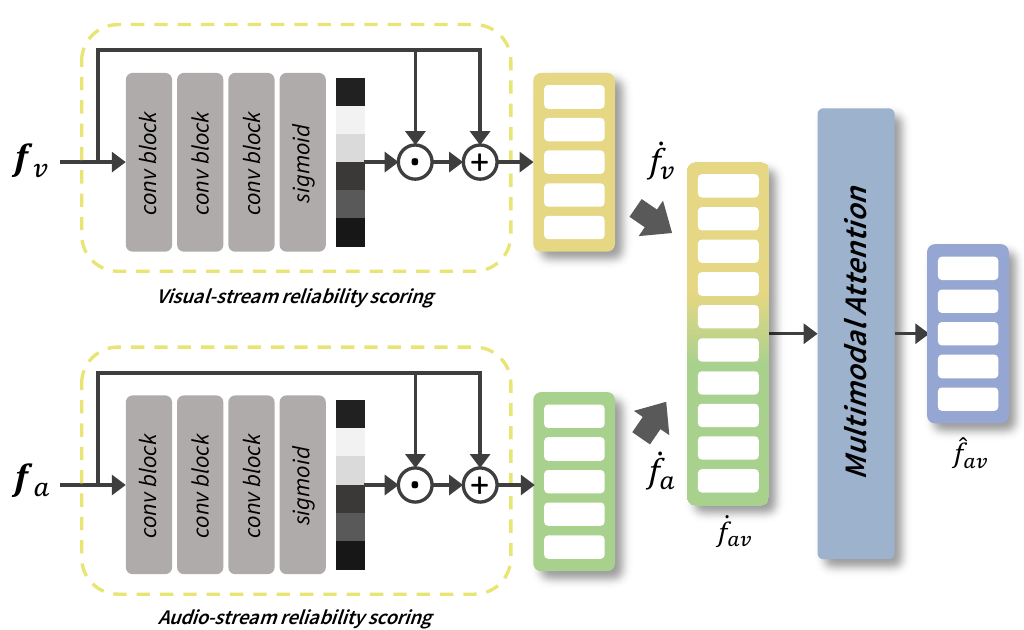}}
	\end{minipage}
	\vspace{-0.5cm}
	\caption{Detailed architecture of AV-RelScore.}
	\label{fig:5}
	\vspace{-0.4cm}
\end{figure}

By using Conformer for the multimodal attentive encoder, the network can attend across modalities so that the reliable modality can be utilized for modeling speech information. Moreover, both intra-modal relationships and inter-modal relationships can be captured through local convolution and global attention. Finally, the first $T$ output features from the multimodal attentive encoder are utilized to predict the sentence via Transformer decoder \cite{vaswani2017attention}. The visualization of the proposed AV-RelScore is shown in Figure \ref{fig:5}.

\subsubsection{Objective functions}
The proposed AVSR framework is trained in an end-to-end manner with audio-visual input corruption. For the objective function, we utilize joint CTC/attention \cite{kim2017joint}. CTC \cite{graves2006ctc} loss is defined as $p_{c}(y|x)\approx \Pi_{t=1}^Tp(y_t|x)$ with an independent assumption of each output, and attention-based loss is defined as $p_{a}(y|x)=\Pi_{j=1}^Jp(y_j|y_{< j},x)$ that the current prediction is determined by previous predictions and inputs, thus including the learning of internal language model, where $J$ represents the total length of ground-truth text. Then, the total objective can be written as follows, $\mathcal{L} = \lambda\log p_a(y|x) + (1-\lambda) \log p_c(y|x)$, where $\lambda$ is a weight parameter for balancing two loss terms.

%------------------------------------ Figure 5
%#################################################
%#################################################
%------------------------------------------------------------------------

\section{Experimental setup}
\label{sec:experiments}

\subsection{Dataset}
\noindent {\bf LRS2} \cite{chung2017lrs2} is an English sentence-level audio-visual dataset collected from BBC television shows. It has about 142,000 utterances including pre-train and train sets, about 1,000 utterances for validation set, and about 1,200 utterances for test set. We utilize both sets for training, and test the model on the test set.

\noindent {\bf LRS3} \cite{afouras2018lrs3} is another large-scale English sentence-level audio-visual dataset. It consists of about 150,000 videos which are a total of about 439 hours long and collected from TED. About 131,000 utterances are utilized for training, and about 1,300 utterances are used for testing.

\subsection{Architecture details}
We adopt the visual front-end and the audio front-end from \cite{ma2021end}. The visual front-end module is comprised of a 3D convolutional layer with a kernel size of $5\times7\times7$ followed by a ResNet18 \cite{he2016deep}. Then the output features are squeezed along the spatial dimension by a global average pooling layer. The audio front-end module consists of a 1D convolutional layer with blocks of ResNet18. Both visual and audio front-ends are initialized using a pre-trained model on LRW \cite{chung2016lrw}.

For the multimodal attention with Conformer encoder \cite{gulati2020conformer}, we use hidden dimensions of 256, feed-forward dimensions of 2048, 12 layers, 8 attention heads, and a convolution kernel size of 31. We utilize Transformer decoder \cite{vaswani2017attention} for prediction, and the decoder is composed of hidden dimensions of 256, feed-forward dimensions of 2048, 6 layers, and 8 attention heads.

For the Audio-Stream Reliability Scoring module and the Visual-Stream Reliability Scoring module, we exploit three 1D convolution layers where each layer is followed by batch normalization and ReLU activation function. The output features are taken into sigmoid activation for obtaining scores in the range of $\left[0, 1\right]$.

%------------------------------------ Figure 6
%#################################################
\begin{figure}[t!]
	\begin{minipage}[b]{1.0\linewidth}
		\centering
		\centerline{\includegraphics[width=8cm]{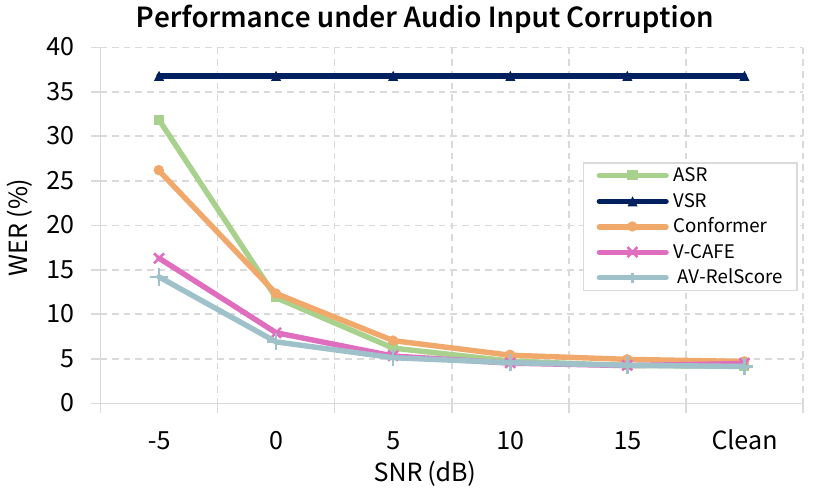}}
	\end{minipage}
	\vspace{-0.7cm}
	\caption{Speech recognition performances of ASR, VSR, and the proposed model under audio input-only corruption using LRS2 dataset. Note the models are trained with audio-visual corruption modeling.}
	\label{fig:6} 
	\vspace{-0.4cm}
\end{figure}
%##################################################

\begin{table*}[]
	\renewcommand{\arraystretch}{1.25}
	\renewcommand{\tabcolsep}{1.1mm}
\centering
\resizebox{0.999999\linewidth}{!}{
\begin{tabular}{ccc cccccc cccccc cccccc} 
\Xhline{3\arrayrulewidth}
\multirow{2.5}{*}{\textbf{Dataset}} & \multirow{2.5}{*}{\makecell{\textbf{Input}\\\textbf{Modal}}}& \multirow{2.5}{*}{\textbf{Method}} & \multicolumn{6}{c}{\textbf{Occlusion}} & \multicolumn{6}{c}{\textbf{Noise}} & \multicolumn{6}{c}{\textbf{Occlusion+Noise}} \\  \cmidrule(l{2pt}r{2pt}){4-9} \cmidrule(l{2pt}r{2pt}){10-15} \cmidrule(l{2pt}r{2pt}){16-21} 
 & & & \textbf{clean}  & \textbf{15}  & \textbf{10}  & \textbf{5} & \textbf{0} & \textbf{-5}  & \textbf{clean}  & \textbf{15}  & \textbf{10}  & \textbf{5} & \textbf{0} & \textbf{-5}  & \textbf{clean}  & \textbf{15}  & \textbf{10}  & \textbf{5} & \textbf{0} & \textbf{-5} \\ \midrule
\multirow{5}{*}{\textbf{LRS2}} 
& A & ASR \cite{ma2021end} & 4.17 & 4.37 & 4.76 & 5.57 & 8.26 & 17.58 & \textbf{4.17} & \textbf{4.37} & 4.76 & 5.57 & 8.26 & 17.58 & \textbf{4.17} & 4.37 & 4.76 & 5.57 & 8.26 & 17.58 \\ 
& V & VSR \cite{ma2021end} & 48.11 & 48.11 & 48.11 & 48.11 & 48.11 & 48.11 & 60.11 & 60.11 & 60.11 & 60.11 & 60.11 & 60.11 & 69.61 & 69.61 & 69.61 & 69.61 & 69.61 & 69.61 \\ \cdashline{2-21}
& A + V & Conformer\cite{ma2021end} & 4.91 & 5.17 & 5.36 & 6.51 & 9.85 & 17.30 & 4.84 & 5.06 & 5.33 & 6.40 & 9.67 & 17.66 & 5.03 & 5.32 & 5.63 & 6.56 & 10.41 & 20.15 \\ 
& A + V & V-CAFE\cite{hong2022vcafe} & 4.57 & 4.44 & 4.84 & 5.50 & 7.34 & 12.15 & 4.87 & 4.64 & 4.93 & 5.44 & 7.12 & \textbf{11.62} & 4.69 & 4.55 & 5.06 & 5.66 & 7.78 & 14.07 \\ \cline{2-21}
& A + V & \textbf{AV-RelScore} & \textbf{4.16} & \textbf{4.34} &  \textbf{4.37} & \textbf{5.21} & \textbf{6.38} & \textbf{11.32} & 4.54 & 4.42 & \textbf{4.45} & \textbf{5.24} & \textbf{6.31} & 11.79 & 4.25 & \textbf{4.35} & \textbf{4.49} & \textbf{5.45} & \textbf{6.95} & \textbf{13.36} \\
\midrule

\multirow{5}{*}{\textbf{LRS3}} 
& A & ASR \cite{ma2021end} & \textbf{2.53} & \textbf{2.68} & 2.97 & 3.53 & 5.64 & 12.95 & \textbf{2.53} & \textbf{2.68} & 2.97 & 3.53 & 5.64 & 12.95& \textbf{2.53} & \textbf{2.68} & \textbf{2.97} & 3.53 & 5.64 & 12.95 \\ 
& V & VSR \cite{ma2021end} & 56.45  & 56.45 & 56.45 & 56.45 & 56.45 & 56.45 & 61.45 & 61.45 & 61.45 & 61.45 & 61.45 & 61.45 & 71.52 & 71.52 & 71.52 & 71.52 & 71.52 & 71.52 \\ \cdashline{2-21}
& A + V & Conformer\cite{ma2021end} & 2.93 & 3.11 & 3.32& 3.79 & 5.61 & 10.98 & 3.00 & 3.00 & 3.32 & 3.79 & 5.62 & 10.62 & 3.03 & 3.03 & 3.33 & 3.85 & 5.64 & 11.82\\
& A + V & V-CAFE\cite{hong2022vcafe} & 3.39 & 3.38 & 3.46 & 3.84 & 5.34 & 9.00 & 3.49 & 3.48 & 3.63 & 3.83 & 5.31 & 8.69 & 3.67 & 3.37 & 3.69 & 4.17 & 5.70 & 10.04 \\ \cline{2-21}
& A + V & \textbf{AV-RelScore} 
& 2.91 & 2.83 & \textbf{2.89} & \textbf{3.25} & \textbf{4.81} & \textbf{8.70} & 3.05 & 2.89 & \textbf{2.92} & \textbf{3.31} & \textbf{4.61} & \textbf{8.51} & 2.95 & 2.91 & 3.10 & \textbf{3.34} & \textbf{5.11} & \textbf{9.41} \\\Xhline{3\arrayrulewidth}
\end{tabular}}
\vspace{-0.2cm}
\caption{WER (\%) comparisons with the state-of-the-art methods on audio-visual corrupted environment. The first row represents the types of visual corruption: patch occlusion, noise, and both, and the second row indicates audio noise with different levels, SNR(dB).}
\label{table:1}
\vspace{-0.4cm}
\end{table*}

\subsection{Implementation details}
For training and testing, every input frame is converted into grayscale. For data augmentation purposes, random cropping and horizontal flipping are applied to the visual inputs during training. For visual corruption modeling, occlusion is modeled with maximum occurrence number (\ie, $N$) as 3, Gaussian blur with a kernel size of 7 and random sigma range of 0.1 to 2.0 is utilized for corruption using blur, and Gaussian noise with a maximum variance of 0.2 is utilized for additive noise corruption. For audio corruption modeling, we set the same setting as \cite{ma2021end} that utilizes a babble noise of \cite{martinez2020mstcn} with an SNR level from -5dB to 20dB. We adopt curriculum learning \cite{bengio2009curriculum}. We initially train the model with the input videos that have lengths within 100 frames. Then, the model is again trained with those with lengths within 150 frames, 300 frames, and finally 600 frames length. We train 50 epochs for 100 and 150 frames, and 20 epochs for 300 and 600 frames. The whole network is trained with Adam optimizer \cite{kingma2014adam} with $\beta_1=0.9$, $\beta_2=0.98$, and $\epsilon=10^{-9}$. We utilize a learning rate scheduler and follow the same scheme as \cite{ma2022visual}, where it increases linearly in the first 25,000 steps, yielding a peak learning rate of 0.0004 and thereafter decreasing in proportional to the inverse square root of the step number. We utilize 4 Nvidia RTX 3090 GPUs for training.

During decoding, we use beam search with a beam width of 40 and an external language model trained on LRS2\cite{chung2017lrs2}, LRS3\cite{afouras2018lrs3}, LibriSpeech \cite{panayotov2015librispeech}, Voxforge, TED-LIUM 3\cite{hernandez2018ted}, and Common Voice \cite{ardila2019common}, following \cite{ma2022visual}. Therefore, the decoding procedure can be written as follows,
\begin{align}
    p(y|x)=\alpha\log p_a(y|x) + (1-\alpha) \log p_c(y|x) \\ + \, \beta \log p_{lm}(y),\notag
\end{align}
where $p_{lm}$ is the decoding score from the external language model, and $\alpha$ and $\beta$ are the weight parameters. We use 0.9 for $\lambda$ in the training objective function, 0.9 and 0.5 for $\alpha$ and $\beta$ in LRS2, respectively, and 0.9 and 0.6 for $\alpha$ and $\beta$ in LRS3, respectively. Note that we re-implement all the previous methods and reproduce the results with the same experimental settings as ours for fair comparisons.

\section{Experimental results}
\subsection{Robustness to audio-visual corruption}
To begin with, we compare the proposed AVSR framework with the state-of-the-art methods including ASR and VSR on LRS2 and LRS3 datasets. We report performances under different corruption environments with three types of visual corruption modeling: occlusion, noise, and occlusion+noise, and audio corruption modeling: an SNR range from -5 to 15dB and a clean-audio environment. Table \ref{table:1} shows the comparison results. The results emphasize the importance of audio-visual corruption modeling. All AVSR models achieve better performances than the ASR model under strong acoustic noise situations of -5, 0, and 5dB SNR even if there is strong visual corruption, the combination of occlusion and noises. By comparing the results with that shown in Figure \ref{fig:3}(c) that the ASR model always yields the best performance under audio-visual corruption, we can confirm that the proposed audio-visual corruption modeling is important in building robust AVSR models. More importantly, when the SNR is lower (-5 to 5dB), meaning that the audio corruption is much applied, and the visual inputs are also corrupted, the proposed model, AV-RelScore outperforms all other previous methods including ASR model. This clearly verifies that our proposed AV-RelScore module effectively finds and utilizes the more reliable modal for recognizing speech.

In addition, we also compare the performances of each method in an audio-only corrupted environment which is the standard setting of previous methods \cite{ma2021end,hong2022vcafe} so that all audio sequences are corrupted with babble noise, instead of corrupting random chunks. Figure \ref{fig:6} shows the comparison results using LRS2 dataset. For this case, AVSR models show the best robustness compared to ASR model. Moreover, the proposed AV-RelScore outperforms the other methods in severe noise conditions.

\begin{table}[]
	\renewcommand{\arraystretch}{1.3}
 	\renewcommand{\tabcolsep}{3mm}
\centering

\resizebox{0.8\linewidth}{!}{
\begin{tabular}{cccc}
\Xhline{3\arrayrulewidth}
\multicolumn{3}{c}{\textbf{Proposed Method}} &\\ \cmidrule{1-3}
Baseline & \makecell{Multimodal\\attention} &\makecell{Reliability\\scoring}  & \textbf{WER(\%)} \\ \hline
\cmark & \xmark & \xmark & 20.15  \\
\cmark & \cmark & \xmark & 13.70 \\
% \cmark &  \cmark  & \xmark &  \\ \hdashline
\cmark & \cmark & \cmark & \textbf{13.36} \\ % << lm0.5

\Xhline{3\arrayrulewidth}
\end{tabular}}
\vspace{-0.1cm}
\caption{Ablation study on LRS2 dataset.}
\label{table:2}
\vspace{-0.5cm}
\end{table}

%------------------------------------ Figure 7
%#################################################
\begin{figure*}[t!]
	\begin{minipage}[b]{1.0\linewidth}
		\centering
		\centerline{\includegraphics[width=15.1cm]{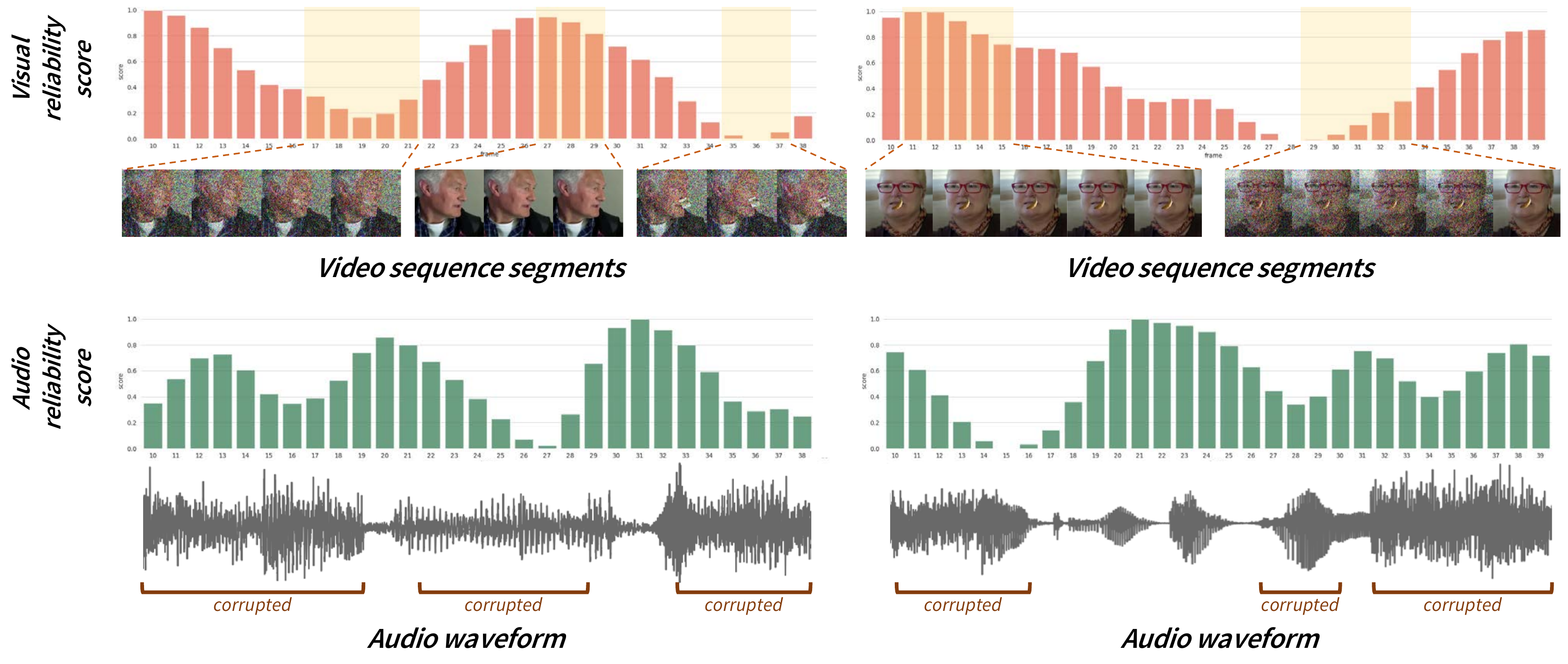}}
	\end{minipage}
	\vspace{-0.7cm}
	\caption{Visualization of visual reliability scores and audio reliability scores from AV-RelScore module of LRS2 dataset.}
	\label{fig:7}
	\vspace{-0.4cm}
\end{figure*}
%##################################################
%------------------------------------------------------------------------

\subsection{Visualization of reliability score}
In this section, we analyze the effectiveness of our proposed AV-RelScore module by visualizing the reliability scores of audio and visual modality, shown in Figure \ref{fig:7}. The visual reliability scores are represented with red bars, and the audio reliability scores are indicated as green bars. From the visual and audio reliability scores, we can clearly observe that the more highly corrupted visual inputs are, the fewer reliability scores are obtained. In addition, if the occlusion patch directly covers the lip movements, shown in the $3^{rd}$ video segment of the first example, the visual reliability score is much less than that with the occlusion patch located slightly left to the lip, shown in the $1^{st}$ video segment of the first example. If the occlusion patch is not covering the lip movements, the visual reliability scores are high enough to recognize the speech properly, shown in the $1^{st}$ video segment of the second example. The scores are gradually increasing when the visual corruption is getting eliminated, indicated in the $2^{nd}$ video segment of the second example. We also notice that each visual reliability and audio reliability score properly supplement each other when there is one modal stream corruption, so when the visual reliability scores are low, the audio reliability scores are high enough to help recognize the speech.

\begin{table}[]
	\renewcommand{\arraystretch}{1.1}
	\renewcommand{\tabcolsep}{5.7mm}
\centering
\resizebox{0.76\linewidth}{!}{\normalsize
\begin{tabular}{ccc}
\Xhline{3\arrayrulewidth}
\textbf{Method}  & \textbf{LRS2} & \textbf{LRS3} \\ \hline
TM-Seq2Seq \cite{afouras2018deep}  & 8.5 & 7.2 \\ 
CTC/Attention \cite{petridis2018avsrhybrid} & 7.0 & - \\ 
LF-MMI TDNN \cite{yu2020overlaplrs2} & 5.9 & -  \\ 
EG-Seq2Seq \cite{xu2020discriminative}  & - & 6.8 \\
RNN-T \cite{makino2019RNNt} & - & 4.5 \\ 
Conformer \cite{ma2021end}& 4.7 & 3.2 \\
V-CAFE\cite{hong2022vcafe}& 4.5 & 3.4 \\\hline
\textbf{AV-RelScore} & \textbf{4.1} & \textbf{2.8} \\
\Xhline{3\arrayrulewidth}
\end{tabular}
}
\vspace{-0.1cm}
\caption{WER (\%) comparisons with state-of-the-art methods.}
\label{table:3}
\vspace{-0.5cm}
\end{table}

\subsection{Ablation study}
\vspace{-0.1cm}
We conduct an ablation study to confirm the effect of the proposed architecture on LRS2. We evaluate the performances in audio-visual corrupted dataset (audio: -5dB SNR, visual: occlusion+noise). To this end, we examine the performances of different variants of the proposed model. By setting the baseline network as \cite{ma2021end} which does not contain both reliability scoring and multimodal attention, we firstly add multimodal attention from the baseline architecture, and the performance improves to 13.70\% WER from 20.15\% WER, indicated in the second row of Table \ref{table:2}. The results show that it is beneficial to use cross-modal attention so that the model can refer to other modalities when the given modality is less informative, compared to the method of independent modeling and fusing the two representations with a linear layer. Then, we add the reliability scoring which is the final proposed model. The performance attains 13.36\% WER. Since it is important to evaluate the significance of each visual frame and audio sequence, especially when both are corrupted, the AV-RelScore is necessary for achieving better performance. From the ablation study, it is clearly implied that it is important to not only refer to other modal features when one modal feature is corrupted but also consider essential parts of the corrupted audio-visual inputs, through both Reliability scoring.

\subsection{Comparisons with audio-visual corruption free}
\vspace{-0.1cm}
We also verify the proposed AV-RelScore by comparing with the previous state-of-the-art methods \cite{afouras2018deep, petridis2018avsrhybrid, yu2020overlaplrs2, xu2020discriminative, xu2020discriminative, makino2019RNNt, ma2021end, hong2022vcafe} in audio-visual clean environment, shown in Table \ref{table:3}. The results indicate that the proposed AVSR framework outperforms the state-of-the-art methods even in a clean environment, by achieving 4.1\% WER and 2.8\% WER on the LRS2 and LRS3 datasets, respectively.

\section{Conclusion}
\label{sec:conclusion}
\vspace{-0.1cm}
In this paper, we propose audio-visual corruption modeling for AVSR and show its importance in boosting the robustness of AVSR systems. Moreover, we propose AV-RelScore that determines which modal input stream is more meaningful than others and emphasizes the meaningful representations. The effectiveness of the proposed audio-visual corruption modeling and AV-RelScore is verified through comprehensive analysis and experiments on two large-scale audio-visual databases, LRS2 and LRS3.
\vspace{0.1cm}

\noindent \textbf{Acknowledgement} This work was partially supported by the National Research Foundation of Korea (NRF) grant funded by the Korea government (MSIT) (No.NRF-2022R1A2C2005529), and Institute of Information $\&$ communications Technology Planning $\&$ Evaluation (IITP) grant funded by the Korea government(MSIT) (No.2022-0-00124, Development of Artificial Intelligence Technology for Self-Improving Competency-Aware Learning Capabilities).

\clearpage
%%%%%%%%% REFERENCES
{\small
\bibliographystyle{ieee_fullname}
\bibliography{egbib}

\begin{thebibliography}{10}\itemsep=-1pt

\bibitem{amodei2016deep2}
Dario Amodei, Sundaram Ananthanarayanan, Rishita Anubhai, Jingliang Bai, Eric
  Battenberg, Carl Case, Jared Casper, Bryan Catanzaro, Qiang Cheng, Guoliang
  Chen, et~al.
\newblock Deep speech 2: End-to-end speech recognition in english and mandarin.
\newblock In {\em International conference on machine learning}, pages
  173--182. PMLR, 2016.

\bibitem{assael2016lipnet}
Yannis~M Assael, Brendan Shillingford, Shimon Whiteson, and Nando De~Freitas.
\newblock Lipnet: End-to-end sentence-level lipreading.
\newblock {\em arXiv preprint arXiv:1611.01599}, 2016.

\bibitem{weninger2015speech}
Felix Weninger, Hakan Erdogan, Shinji Watanabe, Emmanuel Vincent, Jonathan~Le
  Roux, John~R Hershey, and Bj{\"o}rn Schuller.
\newblock Speech enhancement with lstm recurrent neural networks and its
  application to noise-robust asr.
\newblock In {\em International conference on latent variable analysis and
  signal separation}, pages 91--99. Springer, 2015.

\bibitem{tan2019learning}
Ke Tan and DeLiang Wang.
\newblock Learning complex spectral mapping with gated convolutional recurrent
  networks for monaural speech enhancement.
\newblock {\em IEEE/ACM Transactions on Audio, Speech, and Language
  Processing}, 28:380--390, 2019.

\bibitem{wang2020complex}
Zhong-Qiu Wang, Peidong Wang, and DeLiang Wang.
\newblock Complex spectral mapping for single-and multi-channel speech
  enhancement and robust asr.
\newblock {\em IEEE/ACM transactions on audio, speech, and language
  processing}, 28:1778--1787, 2020.

\bibitem{braun2021towards}
Sebastian Braun, Hannes Gamper, Chandan~KA Reddy, and Ivan Tashev.
\newblock Towards efficient models for real-time deep noise suppression.
\newblock In {\em ICASSP 2021-2021 IEEE International Conference on Acoustics,
  Speech and Signal Processing (ICASSP)}, pages 656--660. IEEE, 2021.

\bibitem{noda2015audio}
Kuniaki Noda, Yuki Yamaguchi, Kazuhiro Nakadai, Hiroshi~G Okuno, and Tetsuya
  Ogata.
\newblock Audio-visual speech recognition using deep learning.
\newblock {\em Applied Intelligence}, 42(4):722--737, 2015.

\bibitem{afouras2018deep}
Triantafyllos Afouras, Joon~Son Chung, Andrew Senior, Oriol Vinyals, and Andrew
  Zisserman.
\newblock Deep audio-visual speech recognition.
\newblock {\em IEEE transactions on pattern analysis and machine intelligence},
  2018.

\bibitem{vaswani2017attention}
Ashish Vaswani, Noam Shazeer, Niki Parmar, Jakob Uszkoreit, Llion Jones,
  Aidan~N Gomez, {\L}ukasz Kaiser, and Illia Polosukhin.
\newblock Attention is all you need.
\newblock {\em Advances in neural information processing systems}, 30, 2017.

\bibitem{petridis2018end}
Stavros Petridis, Themos Stafylakis, Pingehuan Ma, Feipeng Cai, Georgios
  Tzimiropoulos, and Maja Pantic.
\newblock End-to-end audiovisual speech recognition.
\newblock In {\em 2018 IEEE international conference on acoustics, speech and
  signal processing (ICASSP)}, pages 6548--6552. IEEE, 2018.

\bibitem{gulati2020conformer}
Anmol Gulati, James Qin, Chung-Cheng Chiu, Niki Parmar, Yu Zhang, Jiahui Yu,
  Wei Han, Shibo Wang, Zhengdong Zhang, Yonghui Wu, et~al.
\newblock Conformer: Convolution-augmented transformer for speech recognition.
\newblock {\em arXiv preprint arXiv:2005.08100}, 2020.

\bibitem{hong2022vcafe}
Joanna Hong, Minsu Kim, Daehun Yoo, and Yong~Man Ro.
\newblock Visual context-driven audio feature enhancement for robust end-to-end
  audio-visual speech recognition.
\newblock {\em arXiv preprint arXiv:2207.06020}, 2022.

\bibitem{wang2019multigrained}
Chenhao Wang.
\newblock Multi-grained spatio-temporal modeling for lip-reading.
\newblock {\em arXiv preprint arXiv:1908.11618}, 2019.

\bibitem{xiao2020deformation}
Jingyun Xiao, Shuang Yang, Yuanhang Zhang, Shiguang Shan, and Xilin Chen.
\newblock Deformation flow based two-stream network for lip reading.
\newblock {\em arXiv preprint arXiv:2003.05709}, 2020.

\bibitem{zhao2020mi}
X. Zhao, S. Yang, S. Shan, and X. Chen.
\newblock Mutual information maximization for effective lip reading.
\newblock In {\em 2020 15th IEEE International Conference on Automatic Face and
  Gesture Recognition (FG 2020) (FG)}, pages 843--850, Los Alamitos, CA, USA,
  may 2020. IEEE Computer Society.

\bibitem{zhang2020cutout}
Yuanhang Zhang, Shuang Yang, Jingyun Xiao, Shiguang Shan, and Xilin Chen.
\newblock Can we read speech beyond the lips? rethinking roi selection for deep
  visual speech recognition.
\newblock {\em arXiv preprint arXiv:2003.03206}, 2020.

\bibitem{Kim_2021_ICCV}
Minsu Kim, Joanna Hong, Se~Jin Park, and Yong~Man Ro.
\newblock Multi-modality associative bridging through memory: Speech sound
  recollected from face video.
\newblock In {\em Proceedings of the IEEE/CVF International Conference on
  Computer Vision (ICCV)}, pages 296--306, October 2021.

\bibitem{kim2021lip}
Minsu Kim, Joanna Hong, and Yong~Man Ro.
\newblock Lip to speech synthesis with visual context attentional gan.
\newblock {\em Advances in Neural Information Processing Systems},
  34:2758--2770, 2021.

\bibitem{kim2023lip}
Minsu Kim, Joanna Hong, and Yong~Man Ro.
\newblock Lip-to-speech synthesis in the wild with multi-task learning.
\newblock {\em arXiv preprint arXiv:2302.08841}, 2023.

\bibitem{hong2021speech}
Joanna Hong, Minsu Kim, Se~Jin Park, and Yong~Man Ro.
\newblock Speech reconstruction with reminiscent sound via visual voice memory.
\newblock {\em IEEE/ACM Transactions on Audio, Speech, and Language
  Processing}, 29:3654--3667, 2021.

\bibitem{hong2022visagesyntalk}
Joanna Hong, Minsu Kim, and Yong~Man Ro.
\newblock Visagesyntalk: Unseen speaker video-to-speech synthesis via
  speech-visage feature selection.
\newblock In {\em Computer Vision--ECCV 2022: 17th European Conference, Tel
  Aviv, Israel, October 23--27, 2022, Proceedings, Part XXXVI}, pages 452--468.
  Springer, 2022.

\bibitem{baevski2020wav2vec2}
Alexei Baevski, Yuhao Zhou, Abdelrahman Mohamed, and Michael Auli.
\newblock wav2vec 2.0: A framework for self-supervised learning of speech
  representations.
\newblock {\em Advances in Neural Information Processing Systems},
  33:12449--12460, 2020.

\bibitem{kim2022distinguishing}
Minsu Kim, Jeong~Hun Yeo, and Yong~Man Ro.
\newblock Distinguishing homophenes using multi-head visual-audio memory for
  lip reading.
\newblock In {\em Proceedings of the 36th AAAI Conference on Artificial
  Intelligence, Vancouver, BC, Canada}, volume~22, 2022.

\bibitem{ma2021end}
Pingchuan Ma, Stavros Petridis, and Maja Pantic.
\newblock End-to-end audio-visual speech recognition with conformers.
\newblock In {\em ICASSP 2021-2021 IEEE International Conference on Acoustics,
  Speech and Signal Processing (ICASSP)}, pages 7613--7617. IEEE, 2021.

\bibitem{chung2017lrs2}
Joon~Son Chung, Andrew Senior, Oriol Vinyals, and Andrew Zisserman.
\newblock Lip reading sentences in the wild.
\newblock In {\em 2017 IEEE Conference on Computer Vision and Pattern
  Recognition (CVPR)}, pages 3444--3453. IEEE, 2017.

\bibitem{afouras2018lrs3}
Triantafyllos Afouras, Joon~Son Chung, and Andrew Zisserman.
\newblock Lrs3-ted: a large-scale dataset for visual speech recognition.
\newblock {\em arXiv preprint arXiv:1809.00496}, 2018.

\bibitem{mohamed2011acoustic}
Abdel-rahman Mohamed, George~E Dahl, and Geoffrey Hinton.
\newblock Acoustic modeling using deep belief networks.
\newblock {\em IEEE transactions on audio, speech, and language processing},
  20(1):14--22, 2011.

\bibitem{hinton2012speech}
Geoffrey Hinton, Li Deng, Dong Yu, George~E. Dahl, Abdel-rahman Mohamed,
  Navdeep Jaitly, Andrew Senior, Vincent Vanhoucke, Patrick Nguyen, Tara~N.
  Sainath, and Brian Kingsbury.
\newblock Deep neural networks for acoustic modeling in speech recognition: The
  shared views of four research groups.
\newblock {\em IEEE Signal Processing Magazine}, 29(6):82--97, 2012.

\bibitem{dahl2011context}
George~E Dahl, Dong Yu, Li Deng, and Alex Acero.
\newblock Context-dependent pre-trained deep neural networks for
  large-vocabulary speech recognition.
\newblock {\em IEEE Transactions on audio, speech, and language processing},
  20(1):30--42, 2011.

\bibitem{seide2011conversational}
Frank Seide, Gang Li, and Dong Yu.
\newblock Conversational speech transcription using context-dependent deep
  neural networks.
\newblock In {\em Twelfth annual conference of the international speech
  communication association}, 2011.

\bibitem{abdel2012applying}
Ossama Abdel-Hamid, Abdel-rahman Mohamed, Hui Jiang, and Gerald Penn.
\newblock Applying convolutional neural networks concepts to hybrid nn-hmm
  model for speech recognition.
\newblock In {\em 2012 IEEE international conference on Acoustics, speech and
  signal processing (ICASSP)}, pages 4277--4280. IEEE, 2012.

\bibitem{abdel2014convolutional}
Ossama Abdel-Hamid, Abdel-rahman Mohamed, Hui Jiang, Li Deng, Gerald Penn, and
  Dong Yu.
\newblock Convolutional neural networks for speech recognition.
\newblock {\em IEEE/ACM Transactions on audio, speech, and language
  processing}, 22(10):1533--1545, 2014.

\bibitem{sainath2013improvements}
Tara~N Sainath, Brian Kingsbury, Abdel-rahman Mohamed, George~E Dahl, George
  Saon, Hagen Soltau, Tomas Beran, Aleksandr~Y Aravkin, and Bhuvana
  Ramabhadran.
\newblock Improvements to deep convolutional neural networks for lvcsr.
\newblock In {\em 2013 IEEE workshop on automatic speech recognition and
  understanding}, pages 315--320. IEEE, 2013.

\bibitem{graves2013speech}
Alex Graves, Abdel-rahman Mohamed, and Geoffrey Hinton.
\newblock Speech recognition with deep recurrent neural networks.
\newblock In {\em 2013 IEEE international conference on acoustics, speech and
  signal processing}, pages 6645--6649. Ieee, 2013.

\bibitem{graves2014towards}
Alex Graves and Navdeep Jaitly.
\newblock Towards end-to-end speech recognition with recurrent neural networks.
\newblock In {\em International conference on machine learning}, pages
  1764--1772. PMLR, 2014.

\bibitem{sak2014long}
Ha{\c{s}}im Sak, Andrew Senior, and Fran{\c{c}}oise Beaufays.
\newblock Long short-term memory based recurrent neural network architectures
  for large vocabulary speech recognition.
\newblock {\em arXiv preprint arXiv:1402.1128}, 2014.

\bibitem{graves2006ctc}
Alex Graves, Santiago Fern{\'a}ndez, Faustino Gomez, and J{\"u}rgen
  Schmidhuber.
\newblock Connectionist temporal classification: labelling unsegmented sequence
  data with recurrent neural networks.
\newblock In {\em Proceedings of the 23rd international conference on Machine
  learning}, pages 369--376, 2006.

\bibitem{sutskever2014sequence}
Ilya Sutskever, Oriol Vinyals, and Quoc~V Le.
\newblock Sequence to sequence learning with neural networks.
\newblock {\em Advances in neural information processing systems}, 27, 2014.

\bibitem{hannun2014deep}
Awni Hannun, Carl Case, Jared Casper, Bryan Catanzaro, Greg Diamos, Erich
  Elsen, Ryan Prenger, Sanjeev Satheesh, Shubho Sengupta, Adam Coates, et~al.
\newblock Deep speech: Scaling up end-to-end speech recognition.
\newblock {\em arXiv preprint arXiv:1412.5567}, 2014.

\bibitem{schneider2019wav2vec}
Steffen Schneider, Alexei Baevski, Ronan Collobert, and Michael Auli.
\newblock wav2vec: Unsupervised pre-training for speech recognition.
\newblock {\em arXiv preprint arXiv:1904.05862}, 2019.

\bibitem{baevski2019vq}
Alexei Baevski, Steffen Schneider, and Michael Auli.
\newblock vq-wav2vec: Self-supervised learning of discrete speech
  representations.
\newblock {\em arXiv preprint arXiv:1910.05453}, 2019.

\bibitem{kim2021cromm}
Minsu Kim, Joanna Hong, Se~Jin Park, and Yong~Man Ro.
\newblock Cromm-vsr: Cross-modal memory augmented visual speech recognition.
\newblock {\em IEEE Transactions on Multimedia}, pages 1--1, 2021.

\bibitem{martinez2020lipreading}
Brais Martinez, Pingchuan Ma, Stavros Petridis, and Maja Pantic.
\newblock Lipreading using temporal convolutional networks.
\newblock In {\em ICASSP 2020-2020 IEEE International Conference on Acoustics,
  Speech and Signal Processing (ICASSP)}, pages 6319--6323. IEEE, 2020.

\bibitem{zhang2019spatio}
Xingxuan Zhang, Feng Cheng, and Shilin Wang.
\newblock Spatio-temporal fusion based convolutional sequence learning for lip
  reading.
\newblock In {\em Proceedings of the IEEE/CVF International Conference on
  Computer Vision}, pages 713--722, 2019.

\bibitem{zhao2020hearing}
Ya Zhao, Rui Xu, Xinchao Wang, Peng Hou, Haihong Tang, and Mingli Song.
\newblock Hearing lips: Improving lip reading by distilling speech recognizers.
\newblock In {\em Proceedings of the AAAI Conference on Artificial
  Intelligence}, volume~34, pages 6917--6924, 2020.

\bibitem{ma2022visual}
Pingchuan Ma, Stavros Petridis, and Maja Pantic.
\newblock Visual speech recognition for multiple languages in the wild.
\newblock {\em arXiv preprint arXiv:2202.13084}, 2022.

\bibitem{kim2022speaker}
Minsu Kim, Hyunjun Kim, and Yong~Man Ro.
\newblock Speaker-adaptive lip reading with user-dependent padding.
\newblock In {\em Computer Vision--ECCV 2022: 17th European Conference, Tel
  Aviv, Israel, October 23--27, 2022, Proceedings, Part XXXVI}, pages 576--593.
  Springer, 2022.

\bibitem{kim2023prompt}
Minsu Kim, Hyung-Il Kim, and Yong~Man Ro.
\newblock Prompt tuning of deep neural networks for speaker-adaptive visual
  speech recognition.
\newblock {\em arXiv preprint arXiv:2302.08102}, 2023.

\bibitem{prajwal2022sub}
KR Prajwal, Triantafyllos Afouras, and Andrew Zisserman.
\newblock Sub-word level lip reading with visual attention.
\newblock In {\em Proceedings of the IEEE/CVF Conference on Computer Vision and
  Pattern Recognition}, pages 5162--5172, 2022.

\bibitem{huang2013audio}
Jing Huang and Brian Kingsbury.
\newblock Audio-visual deep learning for noise robust speech recognition.
\newblock In {\em 2013 IEEE international conference on acoustics, speech and
  signal processing}, pages 7596--7599. IEEE, 2013.

\bibitem{mroueh2015deep}
Youssef Mroueh, Etienne Marcheret, and Vaibhava Goel.
\newblock Deep multimodal learning for audio-visual speech recognition.
\newblock In {\em 2015 IEEE International Conference on Acoustics, Speech and
  Signal Processing (ICASSP)}, pages 2130--2134. IEEE, 2015.

\bibitem{stewart2013robust}
Darryl Stewart, Rowan Seymour, Adrian Pass, and Ji Ming.
\newblock Robust audio-visual speech recognition under noisy audio-video
  conditions.
\newblock {\em IEEE transactions on cybernetics}, 44(2):175--184, 2013.

\bibitem{yu2020audio}
Jianwei Yu, Shi-Xiong Zhang, Jian Wu, Shahram Ghorbani, Bo Wu, Shiyin Kang,
  Shansong Liu, Xunying Liu, Helen Meng, and Dong Yu.
\newblock Audio-visual recognition of overlapped speech for the lrs2 dataset.
\newblock In {\em ICASSP 2020-2020 IEEE International Conference on Acoustics,
  Speech and Signal Processing (ICASSP)}, pages 6984--6988. IEEE, 2020.

\bibitem{petridis2018audio}
Stavros Petridis, Themos Stafylakis, Pingchuan Ma, Georgios Tzimiropoulos, and
  Maja Pantic.
\newblock Audio-visual speech recognition with a hybrid ctc/attention
  architecture.
\newblock In {\em 2018 IEEE Spoken Language Technology Workshop (SLT)}, pages
  513--520. IEEE, 2018.

\bibitem{xu2020discriminative}
Bo Xu, Cheng Lu, Yandong Guo, and Jacob Wang.
\newblock Discriminative multi-modality speech recognition.
\newblock In {\em Proceedings of the IEEE/CVF Conference on Computer Vision and
  Pattern Recognition}, pages 14433--14442, 2020.

\bibitem{cen2021deep}
Feng Cen, Xiaoyu Zhao, Wuzhuang Li, and Guanghui Wang.
\newblock Deep feature augmentation for occluded image classification.
\newblock {\em Pattern Recognition}, 111:107737, 2021.

\bibitem{li2020yolo}
Yongjun Li, Shasha Li, Haohao Du, Lijia Chen, Dongming Zhang, and Yao Li.
\newblock Yolo-acn: Focusing on small target and occluded object detection.
\newblock {\em IEEE Access}, 8:227288--227303, 2020.

\bibitem{li2018occlusion}
Yong Li, Jiabei Zeng, Shiguang Shan, and Xilin Chen.
\newblock Occlusion aware facial expression recognition using cnn with
  attention mechanism.
\newblock {\em IEEE Transactions on Image Processing}, 28(5):2439--2450, 2018.

\bibitem{cen2020classification}
Feng Cen, Xiaoyu Zhao, Wuzhuang Li, and Fanglai Zhu.
\newblock Classification of occluded images for large-scale datasets with
  numerous occlusion patterns.
\newblock {\em IEEE Access}, 8:170883--170897, 2020.

\bibitem{li2018patch}
Yong Li, Jiabei Zeng, Shiguang Shan, and Xilin Chen.
\newblock Patch-gated cnn for occlusion-aware facial expression recognition.
\newblock In {\em 2018 24th International Conference on Pattern Recognition
  (ICPR)}, pages 2209--2214. IEEE, 2018.

\bibitem{voo2022delving}
Kenny~TR Voo, Liming Jiang, and Chen~Change Loy.
\newblock Delving into high-quality synthetic face occlusion segmentation
  datasets.
\newblock In {\em Proceedings of the IEEE/CVF Conference on Computer Vision and
  Pattern Recognition}, pages 4711--4720, 2022.

\bibitem{wang2020robust}
Angtian Wang, Yihong Sun, Adam Kortylewski, and Alan~L Yuille.
\newblock Robust object detection under occlusion with context-aware
  compositionalnets.
\newblock In {\em Proceedings of the IEEE/CVF Conference on Computer Vision and
  Pattern Recognition}, pages 12645--12654, 2020.

\bibitem{varga1993noisex92}
Andrew Varga and Herman~JM Steeneken.
\newblock Assessment for automatic speech recognition: Ii. noisex-92: A
  database and an experiment to study the effect of additive noise on speech
  recognition systems.
\newblock {\em Speech communication}, 12(3):247--251, 1993.

\bibitem{thiemann2013demand}
Joachim Thiemann, Nobutaka Ito, and Emmanuel Vincent.
\newblock The diverse environments multi-channel acoustic noise database
  (demand): A database of multichannel environmental noise recordings.
\newblock In {\em Proceedings of Meetings on Acoustics ICA2013}, volume~19,
  page 035081. Acoustical Society of America, 2013.

\bibitem{ko2015audio}
Tom Ko, Vijayaditya Peddinti, Daniel Povey, and Sanjeev Khudanpur.
\newblock Audio augmentation for speech recognition.
\newblock In {\em Sixteenth annual conference of the international speech
  communication association}, 2015.

\bibitem{chen2020uniter}
Yen-Chun Chen, Linjie Li, Licheng Yu, Ahmed El~Kholy, Faisal Ahmed, Zhe Gan, Yu
  Cheng, and Jingjing Liu.
\newblock Uniter: Universal image-text representation learning.
\newblock In {\em European conference on computer vision}, pages 104--120.
  Springer, 2020.

\bibitem{kim2017joint}
Suyoun Kim, Takaaki Hori, and Shinji Watanabe.
\newblock Joint ctc-attention based end-to-end speech recognition using
  multi-task learning.
\newblock In {\em 2017 IEEE international conference on acoustics, speech and
  signal processing (ICASSP)}, pages 4835--4839. IEEE, 2017.

\bibitem{he2016deep}
Kaiming He, Xiangyu Zhang, Shaoqing Ren, and Jian Sun.
\newblock Deep residual learning for image recognition.
\newblock In {\em Proceedings of the IEEE conference on computer vision and
  pattern recognition}, pages 770--778, 2016.

\bibitem{chung2016lrw}
Joon~Son Chung and Andrew Zisserman.
\newblock Lip reading in the wild.
\newblock In {\em Asian Conference on Computer Vision}, pages 87--103.
  Springer, 2016.

\bibitem{martinez2020mstcn}
Brais Martinez, Pingchuan Ma, Stavros Petridis, and Maja Pantic.
\newblock Lipreading using temporal convolutional networks.
\newblock In {\em ICASSP 2020-2020 IEEE International Conference on Acoustics,
  Speech and Signal Processing (ICASSP)}, pages 6319--6323. IEEE, 2020.

\bibitem{bengio2009curriculum}
Yoshua Bengio, J{\'e}r{\^o}me Louradour, Ronan Collobert, and Jason Weston.
\newblock Curriculum learning.
\newblock In {\em Proceedings of the 26th annual international conference on
  machine learning}, pages 41--48, 2009.

\bibitem{kingma2014adam}
Diederik~P Kingma and Jimmy Ba.
\newblock Adam: A method for stochastic optimization.
\newblock {\em arXiv preprint arXiv:1412.6980}, 2014.

\bibitem{panayotov2015librispeech}
Vassil Panayotov, Guoguo Chen, Daniel Povey, and Sanjeev Khudanpur.
\newblock Librispeech: an asr corpus based on public domain audio books.
\newblock In {\em 2015 IEEE international conference on acoustics, speech and
  signal processing (ICASSP)}, pages 5206--5210. IEEE, 2015.

\bibitem{hernandez2018ted}
Fran{\c{c}}ois Hernandez, Vincent Nguyen, Sahar Ghannay, Natalia Tomashenko,
  and Yannick Esteve.
\newblock Ted-lium 3: twice as much data and corpus repartition for experiments
  on speaker adaptation.
\newblock In {\em International conference on speech and computer}, pages
  198--208. Springer, 2018.

\bibitem{ardila2019common}
Rosana Ardila, Megan Branson, Kelly Davis, Michael Henretty, Michael Kohler,
  Josh Meyer, Reuben Morais, Lindsay Saunders, Francis~M Tyers, and Gregor
  Weber.
\newblock Common voice: A massively-multilingual speech corpus.
\newblock {\em arXiv preprint arXiv:1912.06670}, 2019.

\bibitem{petridis2018avsrhybrid}
Stavros Petridis, Themos Stafylakis, Pingchuan Ma, Georgios Tzimiropoulos, and
  Maja Pantic.
\newblock Audio-visual speech recognition with a hybrid ctc/attention
  architecture.
\newblock In {\em 2018 IEEE Spoken Language Technology Workshop (SLT)}, pages
  513--520. IEEE, 2018.

\bibitem{yu2020overlaplrs2}
Jianwei Yu, Shi-Xiong Zhang, Jian Wu, Shahram Ghorbani, Bo Wu, Shiyin Kang,
  Shansong Liu, Xunying Liu, Helen Meng, and Dong Yu.
\newblock Audio-visual recognition of overlapped speech for the lrs2 dataset.
\newblock In {\em ICASSP 2020-2020 IEEE International Conference on Acoustics,
  Speech and Signal Processing (ICASSP)}, pages 6984--6988. IEEE, 2020.

\bibitem{makino2019RNNt}
Takaki Makino, Hank Liao, Yannis Assael, Brendan Shillingford, Basilio Garcia,
  Otavio Braga, and Olivier Siohan.
\newblock Recurrent neural network transducer for audio-visual speech
  recognition.
\newblock In {\em 2019 IEEE automatic speech recognition and understanding
  workshop (ASRU)}, pages 905--912. IEEE, 2019.

\end{thebibliography}
}

\end{document}